\documentclass[a4paper]{jpconf}
\usepackage{graphicx}
\usepackage[pass]{geometry}
\usepackage{graphicx}
\usepackage{amsmath}
\usepackage[utf8]{inputenc}
\usepackage{amsmath}
\usepackage{array}
\usepackage{amsfonts}
\usepackage{epsf}
\usepackage{epsfig}
\usepackage{amssymb}
\usepackage{bbm}
\usepackage{delarray}
\usepackage{subfig}

\begin{document}
\title{Triangular quantum profiles: transmission probability and energy spectrum}

\author{Altuğ Arda}

\address{Department of
Physics Education, Hacettepe University, 06800, Ankara, Turkey}

\ead{arda@hacettepe.edu.tr}

\begin{abstract}
Analytical expressions for the transition probability and the
energy spectrum of the $1D$ Schrödinger equation with position
dependent mass are presented for the triangular quantum barrier
and quantum well. The transmission coefficient is obtained by
using the wave functions written in terms of the Airy's functions
and of the solutions of the Kummer's differential equation. In
order to show the validity of our analyze, an example by taking
some numerical values for GaAs heterostructure is presented.
\end{abstract}

\section{Introduction}
Quantum well [1, 2], quantum dot [3, 4] and quantum wire [5, 6]
heterostructures which are classified as low-dimensional
semiconductor quantum systems have become an important part within
the semiconductor studies. The theoretical and practical
investigation gives some possibilities to produce high quality
quantum heterostructures. The materials consisting of GaAs/AlAs
have been widely used in investigation of the above
heterostructures. However, new types of the materials have been
also taken into account, for example, silicon carbide (SiC) [7].

To study the semiconductor heterostructures within the theoretical
and practical frameworks could give some clues that may be helpful
to advance the semiconductor technology such as development of
some optoelectronic and communication devices based on the quantum
mechanical tunnelling, some visible lasers based on the electronic
spectrum of double quantum well [8] and quantum cascade lasers
based on the electronic transition between levels of the
conduction band [9]. Among the other quantum heterostructures, the
triangular quantum well(s) is also important systems since the
absorption coefficient value is reduced in the experimental
measurement of the electroabsorption when triangular quantum wells
used [10]. Moreover, the current absorption spectra of the
triangular quantum profile could operate with lower driving
voltages [11].

The quantum mechanical tunnelling is a fundamental subject within
the studying of the semiconductor quantum systems. This is so
because studying of tunnelling helps to understand the physical
properties of the related system and gives some hints about the
lasing in quantum well lasers and electron transport in some
devices [12]. According to the above points, it could be
interesting to find the transmission coefficient for the
triangular quantum barrier including a numerical presentation and
to study the bound states of the triangular quantum well for the
$1D$ Schrödinger equation. We find that the transmission
probability oscillates within the range of energy but this
behavior is very slightly. The numerical computation is presented
for the triangular quantum barrier made of GaAs heterostructure
where the thickness is $a=7$ nm while it's maximum value is
$V_{0}=450$ meV. It is observed that our numerical results are in
agreement with the ones obtained for the constant mass case [13].
We point out that the energy levels of the triangular quantum well
is finite and the potential parameter $\alpha$ controls the bound
state numbers. In computation, we take the values of the mass
parameters as $0.067m_{0}$ for GaAs where $m_{0}$ is the free
electron mass.

In the present work, we study the transmission probability of the
$1D$ Schrödinger equation for the triangular potential barrier by
writing the related wave functions in terms of the Airy's
functions and in terms of the solutions of the Kummer's
differential equation, namely, the confluent hypergeometric
functions of the first kind. Among the approaches and methods used
to study the quantum heterostructures [12], our formalism is based
on solving the Schrödinger equation coming from the Hamiltonian
written for the case of position dependent mass [14]. In this
case, the generalization of the standard Hamiltonian is not
trivial because the linear momentum and mass operator no longer
commute. We propose a mass function depending on spatially
coordinate to solve 1D effective Schrödinger equation. Our
formalism includes also an approach meaning that we ignore the
terms including the derivatives of mass in 1D effective
Schrödinger equation by assuming one of the mass parameters goes
to zero.

\section{Analytical expressions}

The generalized Hamiltonian for the case where the mass depends on
the spatially coordinate is given [14]
\begin{eqnarray}
\mathcal{H}=\frac{1}{2}\left[p\,\frac{1}{m}\,p\right]+V\,,
\end{eqnarray}
where $p$ is the linear momentum operator and $V$ is the operator
which defines the potential function. The $1D$ Schrödinger
equation for the position dependent mass obtained from the above
Hamiltonian is written as [14]
\begin{eqnarray}
\left\{\frac{d^2}{dx^2}-\frac{dm(x)/dx}{m(x)}\,\frac{d}{dx}+Hm(x)[E-V(x)]\right\}\phi(x)=0\,,
\end{eqnarray}
where $H=2/\hbar^2$.

We tend to parameterize the mass as
\begin{eqnarray}
m(x)=M_{0}-M_{1}x\,,
\end{eqnarray}
where $M_{0}$ and $M_{1}$ are the arbitrary parameters. For the
rest of the computation, we assume that the terms including the
derivatives of the mass could be ignored when the mass parameter
$M_{1} \rightarrow 0$.

\textbf{2.1. Transmission probability}

The triangular quantum barrier is defined as (Figure 1a) [13]
\begin{eqnarray}
V(x)= \left\{ \begin{array}{lcl} 0 & \mbox{for} & x<0 \\
V_{0}-\alpha x & \mbox{for} & 0<x<a \\ 0 & \mbox{for} & x>a
\end{array}\right.
\end{eqnarray}
where $V_{0}$ represents the maximum value of the potential
profile and $\alpha$ controls the thickness of the quantum
barrier.

Inserting Eqs. (3) and (4) into Eq. (2), taking into account the
above assuming and using a new variable
$y=\left(HEM_{1}\right)^{1/3}x$, we obtain the following equation
\begin{eqnarray}
\left[\frac{d^2}{dy^2}-y\right]\phi_{I}(y)=0\,
\end{eqnarray}
solutions are expressed in terms of the Airy's functions. So we
write the solution for the region I ($x<0$) as [15]
\begin{eqnarray}
\phi_{I}(y)=b_{1}Ai(y)+b_{2}Bi(y)\,.
\end{eqnarray}

Similarly, we obtain the following equation for the region II
($0<x<a$)
\begin{eqnarray}
\left\{\frac{d^2}{dx^2}-[a_{1}x^2+a_{2}x+a_{3}]\right\}\phi_{II}(x)=0\,,
\end{eqnarray}
which can be written by using a new variable
$y=x+\frac{a_{2}}{2a_{1}}$ as
\begin{eqnarray}
\left\{\frac{d^2}{dy^2}-[a_{1}y^2+A_{I}]\right\}\phi_{II}(y)=0\,,
\end{eqnarray}
where
\begin{subequations}
\begin{align}
a_{1}&=HM_{1}\alpha\,,\\
a_{2}&=-H[M_{0}\alpha+M_{1}(V_{0}-E)]\,,\\
a_{3}&=HM_{0}(E-V_{0})\,.
\end{align}
\end{subequations}
and $A_{I}=\frac{4a_{1}a_{3}-a^2_{2}}{4a_{1}}$.

In order to get a more suitable form of Eq. (8), we use a new
variable $z=\sqrt{a_{1}\,}y^2$ and a trial wave function in terms
of $z$ as $\phi_{II}(y)=e^{-z/2}f(z)$ which gives
\begin{eqnarray}
\left\{z\frac{d^2}{dz^2}-\bigl(z-\frac{1}{2}\bigr)\,\frac{d}{dz}-\frac{1}{4}\,\left[1+\frac{A_{I}}{4a_{1}}\right]
\right\}f(z)=0,,
\end{eqnarray}
which is a kind of the Kummer's differential equation having the
form [15]
\begin{eqnarray}
x\frac{d^2y(x)}{dx^2}+(c-x)\frac{dy(x)}{dx}-by(x)=0\,.
\end{eqnarray}
Two linear independent solutions of the above equation is written
as [15]
\begin{eqnarray}
y(x) \sim \,_{1}F_{1}(b;c;x)+U(b;c;x)\,,
\end{eqnarray}
where $\,_{1}F_{1}(b;c;x)$ is the confluent hypergeometric
functions of the first kind and the second part is given
\begin{eqnarray}
U(b;c;x)=\pi\csc(\pi
c)\left[\frac{\,_{1}\bar{F}_{1}(b;c;x)}{\Gamma(b-c+1)}-x^{1-c}\,\frac{1}{\Gamma(b)}\,
_{1}\bar{F}_{1}(b-c+1;2-c;x)\right]\,.
\end{eqnarray}
where $\,_{1}\bar{F}_{1}(b;c;x)$ is the regularized confluent
hypergeometric functions of the first kind [15].

With the help of Eq. (12) we write the solution for the region II
as
\begin{eqnarray}
f(z)=b_{3}\,_{1}F_{1}(\frac{1}{4}\,[1+\frac{A_{I}}{4a_{1}}];\frac{1}{2};z)+b_{4}
U(\frac{1}{4}\,[1+\frac{A_{I}}{4a_{1}}];\frac{1}{2};z)\,,
\end{eqnarray}

Following the same steps for region I, we obtain the following
equation
\begin{eqnarray}
\left[\frac{d^2}{dy^2}-y\right]\phi_{III}(y)=0\,
\end{eqnarray}
where we define a new variable as
$y=\left(HEM_{1}\right)^{1/3}x-\frac{HEM_{0}}{\left(HEM_{1}\right)^{2/3}}$.
We write the physical solution for the region III as
\begin{eqnarray}
\phi_{III}(y)=b_{5}Ai(y)\,,
\end{eqnarray}
where used the properties given as $\lim_{x \to +\infty}Bi(x)\to
\infty$, $\lim_{x \to -\infty}Bi(x)\to 0$ and
$Bi(0)=\frac{1}{\sqrt[6]{3}\,\Gamma(\frac{2}{3})}$ [15].

Using the continuity conditions for $\phi(x)$ and $d\phi(x)/dx$ at
$x=0$ and at $x=a$ and after straightforward calculations we
obtain the following expressions
\begin{subequations}
\begin{align}
&b_{1}Ai(y_{1})+b_{2}Bi(y_{1})=b_{3}f'_{1}+b_{4}f_{7}\,,\\
&\left(HEM_{1}\right)^{1/3}\left[b_{1}Ai'(y_{1})+b_{2}Bi'(y_{1})\right]=b_{3}f_{8}+b_{4}f_{9}\,,\\
&b_{5}Ai(y_{3})=b_{3}g'_{1}+b_{4}g_{7}\,,\\
&\left(HEM_{1}\right)^{1/3}b_{5}Ai'(y_{3})=b_{3}g_{8}+b_{4}g_{9}\,.
\end{align}
\end{subequations}
where prime in $Ai(y)$ and $Bi(y)$ denote derivatives in the above
expressions and the following abbreviations are used
\begin{subequations}
\begin{align}
f_{1}&=\sqrt{a_{1}\,}y_{2}e^{-y^2_{2}\sqrt{a_{1}\,}/2}
\,_{1}F_{1}(\frac{1}{4}\,[1+\frac{A_{I}}{\sqrt{a_{1}\,}}];\frac{1}{2};\sqrt{a_{1}\,}y^2_{2})\,,\\
f_{2}&=\sqrt{a_{1}\,}y_{2}e^{-y^2_{2}\sqrt{a_{1}\,}/2}
\,_{1}F_{1}(\frac{1}{4}\,[5+\frac{A_{I}}{\sqrt{a_{1}\,}}];\frac{3}{2};\sqrt{a_{1}\,}y^2_{2})\,,\\
f_{3}&=\pi\csc\frac{\pi}{2}\sqrt{a_{1}\,}y_{2}\frac{e^{-y^2_{2}\sqrt{a_{1}\,}/2}}{\Gamma(\frac{3}{4}
+\frac{A_{I}}{4\sqrt{a_{1}\,}})}
\,_{1}\bar{F}_{1}(\frac{1}{4}\,[1+\frac{A_{I}}{\sqrt{a_{1}\,}}];\frac{1}{2};\sqrt{a_{1}\,}y^2_{2})\,,\\
f_{4}&=\pi\csc\frac{\pi}{2}\sqrt{a_{1}\,}y_{2}\frac{e^{-y^2_{2}\sqrt{a_{1}\,}/2}}
{2\Gamma(\frac{3}{4}+\frac{A_{I}}{4\sqrt{a_{1}\,}})}
\bigl(1+\frac{A_{I}}{\sqrt{a_{1}\,}}\bigr)
\,_{1}\bar{F}_{1}(\frac{1}{4}\,[5+\frac{A_{I}}{\sqrt{a_{1}\,}}];\frac{3}{2};\sqrt{a_{1}\,}y^2_{2})\,.
\end{align}
\end{subequations}
\begin{subequations}
\begin{align}
f_{5}&=\pi\csc\frac{\pi}{2}\sqrt{a_{1}\,}y_{2}\frac{e^{-y^2_{2}\sqrt{a_{1}\,}/2}}
{\Gamma(\frac{1}{4}+\frac{A_{I}}{4\sqrt{a_{1}\,}})}
\,_{1}\bar{F}_{1}(\frac{1}{4}\,[3+\frac{A_{I}}{\sqrt{a_{1}\,}}];\frac{3}{2};\sqrt{a_{1}\,}y^2_{2})\,,\\
f_{6}&=\pi\csc\frac{\pi}{2}\sqrt{a_{1}\,}y_{2}\frac{e^{-y^2_{2}\sqrt{a_{1}\,}/2}}
{2\Gamma(\frac{1}{4}+\frac{A_{I}}{\sqrt{a_{1}\,}})}
\bigl(3+\frac{A_{I}}{\sqrt{a_{1}\,}}\bigr)
\,_{1}\bar{F}_{1}(\frac{1}{4}\,[7+\frac{A_{I}}{\sqrt{a_{1}\,}}];\frac{5}{2};\sqrt{a_{1}\,}y^2_{2})\,,\\
f'_{1}&=\frac{f_{1}}{\sqrt{a_{1}\,}y_{2}}\,,\\
f'_{3}&=\frac{f_{3}}{\sqrt{a_{1}\,}y_{2}}\,,\\
f'_{5}&=\frac{f_{5}}{a^{1/4}_{1}y_{2}}\,,\\
f_{7}&=f'_{3}-f'_{5}\,,\\
f_{8}&=f_{2}-f_{1}\,,\\
f_{9}&=f_{4}+f_{5}-f_{3}-f_{6}\,.
\end{align}
\end{subequations}
and
\begin{eqnarray}
&&g_{1}=f_{1}(y_{2}\rightarrow y_{4});g_{2}=f_{2}(y_{2}\rightarrow
y_{4});g_{3}=f_{3}(y_{2}\rightarrow
y_{4});g_{4}=f_{4}(y_{2}\rightarrow y_{4})\,,\nonumber\\
&&g_{5}=f_{5}(y_{2}\rightarrow y_{4});g_{6}=f_{6}(y_{2}\rightarrow
y_{4});g'_{1}=f'_{1}(y_{2}\rightarrow
y_{4});g'_{3}=f'_{3}(y_{2}\rightarrow
y_{4})\,,\nonumber\\&&g'_{5}=f'_{5}(y_{2}\rightarrow y_{4});g_{7}=g'_{3}-g'_{5};g_{8}=g_{2}-g_{1};g_{9}=g_{4}+g_{5}-g_{3}-g_{6}\,.
\end{eqnarray}
where we use
$\frac{d}{dz}\,_{1}F_{1}(b;c;z)=\frac{b}{c}\,_{1}F_{1}(b+1;c+1;z)$
and
$\frac{d}{dz}\,_{1}\bar{F}_{1}(b;c;z)=b\,_{1}\bar{F}_{1}(b+1;c+1;z)$
to obtain the above expressions [15]. The arguments of the above
functions coming from the continuity conditions at $x=0$ and $x=a$
are given as
\begin{subequations}
\begin{align}
y_{1}&=-\frac{HEM_{0}}{\left(HEM_{1}\right)^{2/3}}\,,\\
y_{2}&=\frac{a_{2}}{2a_{1}}\,,\\
y_{3}&=\left(HEM_{1}\right)^{1/3}a-\frac{HEM_{0}}{\left(HEM_{1}\right)^{2/3}}\,,\\
y_{4}&=a+\frac{a_{2}}{2a_{1}}\,.
\end{align}
\end{subequations}

With the help of Eqs. (18)-(20), Eqs. (17a)-(17d) gives us the
transmission probability as
\begin{eqnarray}
T=\left|\frac{t_{1}}{t_{2}}\right|^2\,,
\end{eqnarray}
where
\begin{subequations}
\begin{align}
t_{1}&=\frac{1}{\pi}\left(HEM_{1}\right)^{1/3}\left[g'_{1}g_{9}-g_{8}g_{7}\right]\,,\\
t_{2}&=\left[g_{9}Ai(y_{3})-\left(HEM_{1}\right)^{1/3}g_{7}Ai'(y_{3})\right]
\left[\left(HEM_{1}\right)^{1/3}f'_{1}Bi'(y_{1})-f_{8}Bi(y_{1})\right]\nonumber\\
&\times\left[\left(HEM_{1}\right)^{1/3}g'_{1}Ai'(y_{3})-g_{8}Ai(y_{3})\right]
\left[\left(HEM_{1}\right)^{1/3}f_{7}Bi'(y_{1})-f_{9}Bi(y_{1})\right]\,.
\end{align}
\end{subequations}
where the following property of the Airy's functions
$Ai(y)Bi'(y)-Ai'(y)Bi(y)=\frac{1}{\pi}$ [15] is used.

To check the validity of our formalism, we compute the
transmission coefficient in Eq. (22) numerically. For this aim,
the parameters are used: $V_{0}=450$ meV, $a=7$ nm [16], the mass
made of GaAs $M_{0}=0.067m_{0}$ where $m_{0}$ is the free electron
mass $9.1 \times 10^{-31}$ kg [12], $\hbar=1.05 \times 10^{-34}$
J.s and $M_{1}=M_{0}$. Figure (2) shows the dependence of the
transmission probability on the energy of the incident particle.
It is seen that the transmission probability is very slightly
within the energy range and observed that goes to one while the
energy increases. We plot the dependence of the transmission
coefficient on the height and width of quantum barrier in Figures
(3) and (4), respectively. The parameter values of mass are the
same with the ones used in Figure (2) but Figures (3) and (4) are
plotted for $E=0.1$ eV. In both of figures, the transmission
probability is exactly one for initial values of $V_{0}$ and $a$,
respectively. It's value decreases while the values of height and
width of quantum barrier increase. The fluctuations of
transmission coefficient are very small as observed in Figure (2).
Figure (5) shows varying of the tunnelling coefficient according to
the energy of incident particle. It is observed that the
tunnelling coefficient is also very slightly within the energy
range as in Figure (2).

\textbf{2.2. Bound states}

In order to study the bound states of the triangular quantum well,
we parameterize the potential profile as (Figure 1b)
\begin{eqnarray}
V(x)= \left\{ \begin{array}{lcl} 0 & \mbox{for} & x<0 \\
-V_{0}-\alpha x & \mbox{for} & 0<x<a \\ 0 & \mbox{for} & x>a
\end{array}\right.
\end{eqnarray}
In this case, we obtain a differential equation similar to the one
given in Eq. (10) with the following abbreviations
\begin{subequations}
\begin{align}
a'_{1}&=HM_{1}\alpha=a_{1}\,,\\
a'_{2}&=-H[M_{0}\alpha-M_{1}(V_{0}+E)]\,,\\
a'_{3}&=HM_{0}(E+V_{0})\,.
\end{align}
\end{subequations}
and $B_{I}=\frac{4a'_{1}a'_{3}-a'^2_{2}}{4a'_{1}}$ which give a
physical acceptable solution as
\begin{eqnarray}
f(z)
\sim\,_{1}F_{1}(\frac{1}{4}\,[1+\frac{B_{I}}{4a'_{1}}];\frac{1}{2};z)\,.
\end{eqnarray}
In order to get a finite solution it must be
\begin{eqnarray}
\frac{1}{4}\,\bigl(1+\frac{B_{I}}{4a'_{1}}\bigr)=-n\,\, (n\in
\mathbb{N})\,,
\end{eqnarray}
which is a quantization condition for the bound states. We write
the energy spectra of the triangular quantum well as
\begin{eqnarray}
E_{n}=-V_{0}-\frac{M_{0}\alpha}{M_{1}}+2\bigl(\frac{\alpha^{3}}{M_{1}}\bigr)^{1/4}\sqrt{1+4n\,}\,.
\end{eqnarray}
It is worth to say that the parameter $\alpha$ in the potential
profile depends on the diffusion length (or the Debye length)
$L_{D}$ inversely which controls the number of the bound states
[16]. In order to get some numerical values for the bound states
we choose the parameter set as $M_{1}=M_{0}=0.067m_{0}$ and
$\alpha=0.01V_{0}$ and present our results in Table 1. It shows that the results are agreement  with the ones
stated in literature [16].

As a final remark, we tend to discuss briefly the effect of an external electric field on a quantum system considered here. For this case, the potential is written as an 'effective' potential as following
\begin{eqnarray}
V_{eff}=V(x)+V_{E}(x)\,,
\end{eqnarray}
where $V_{E}(x)$ denotes the potential part arising from the external electric field. The second term in Eq. (29) depends on the electric field strength linearly [8]. So we expect that the obtained expressions for transmission coefficient and also energy levels have additional terms including the electric field strength.

\section{Conclusion}
In the present work, we have analyzed the transmission probability
for the $1D$ Schrödinger equation with position-dependent mass for
the triangular quantum barrier by applying the continuity
conditions on the wave functions which are written in terms of the
Airy's functions and the solutions of the Kummer's equation. For
this aim, we have solved the Schrödinger equation obtained from a
non-standard Hamiltonian written for the case where the mass and
linear momentum operator does not commute. We have ignored the
terms including the derivatives of mass in Schrödinger equation
for $M_{1} \rightarrow 0$ to obtain the analytical solutions. We
have given the dependence of the transmission probability not only
on the energy but also on the height and width of the barrier,
respectively. We have observed that the transmission coefficient
decreases while the values of parameters $E$, $V_{0}$ and $a$
increase, as expected, and the fluctuations of the transmission
probability are very small. We have also studied the bound states
of the triangular quantum well and observed that the number of the
bound states is finite depending on the mass parameter $M_{1}$ and
on the diffusion length.

\newpage

\begin{table}
\centering
\caption{Some energy eigenvalues for the triangular quantum profile
(eV).}
\begin{tabular}{ccccc}
 & & \text{Ref. [16]} && \text{our results}  \\ \hline
$E_{1}$ & & -0.20986 && -0.29407 \\
$E_{2}$ & & -0.00630 && -0.00871 \\
\end{tabular}
\end{table}

\smallskip

\newpage

\begin{figure} \centering \subfloat[]{
\includegraphics[height=3in,width=5in]{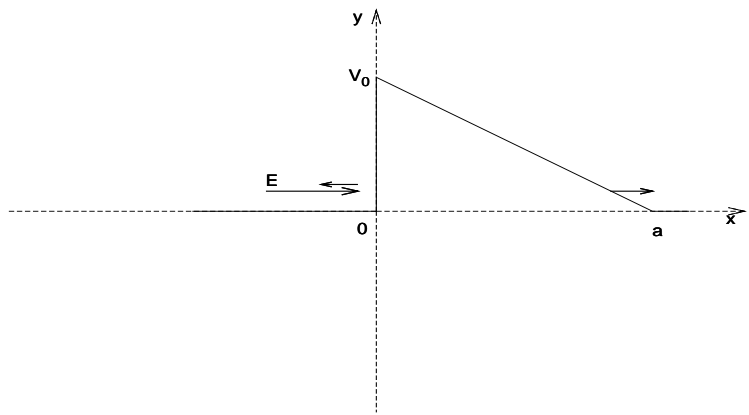}}
\hspace{0.1\linewidth} \subfloat[]{
\includegraphics[height=3in,width=5in]{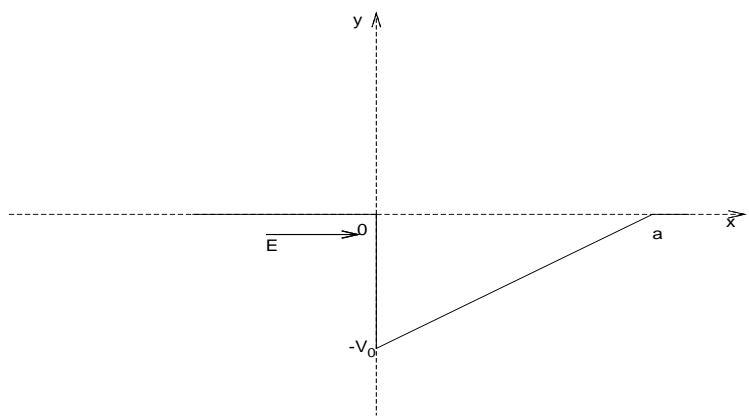}}
\caption{Graphical representation of the triangular quantum
barrier (a) and triangular quantum well (b).}
\end{figure}

\newpage

\begin{figure}
\centering
\includegraphics[height=3in, width=5in, angle=0]{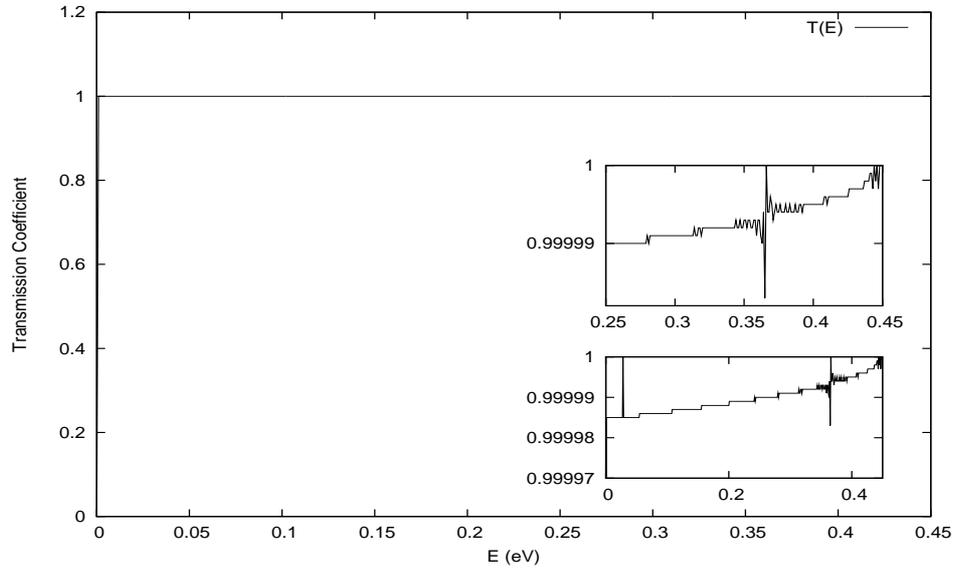}
\caption{The transmission probability for the triangular quantum
barrier versus $E$.}
\end{figure}

\newpage

\begin{figure}
\centering
\includegraphics[height=3in, width=5in, angle=0]{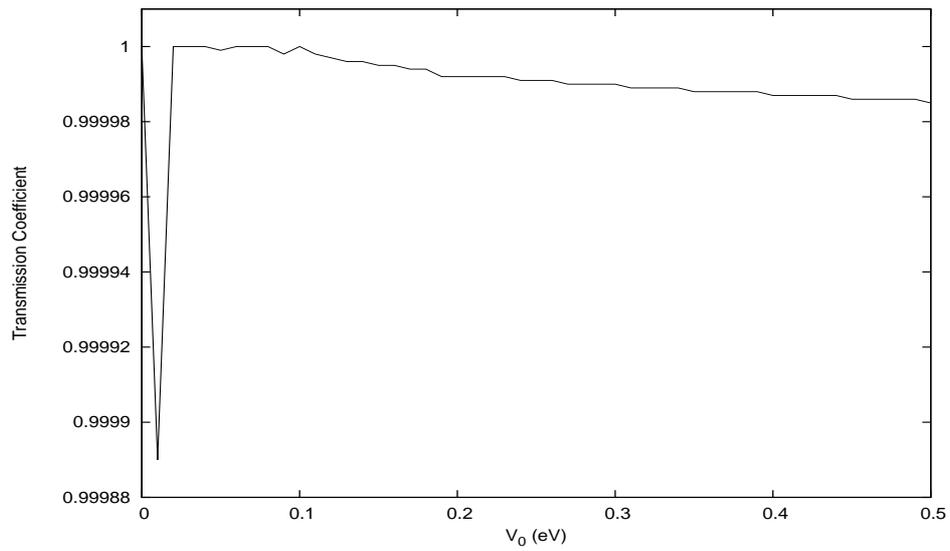}
\caption{The variation of transmission probability versus
 height of barrier $V_{0}$.}
\end{figure}

\newpage

\begin{figure}
\centering
\includegraphics[height=3in, width=5in, angle=0]{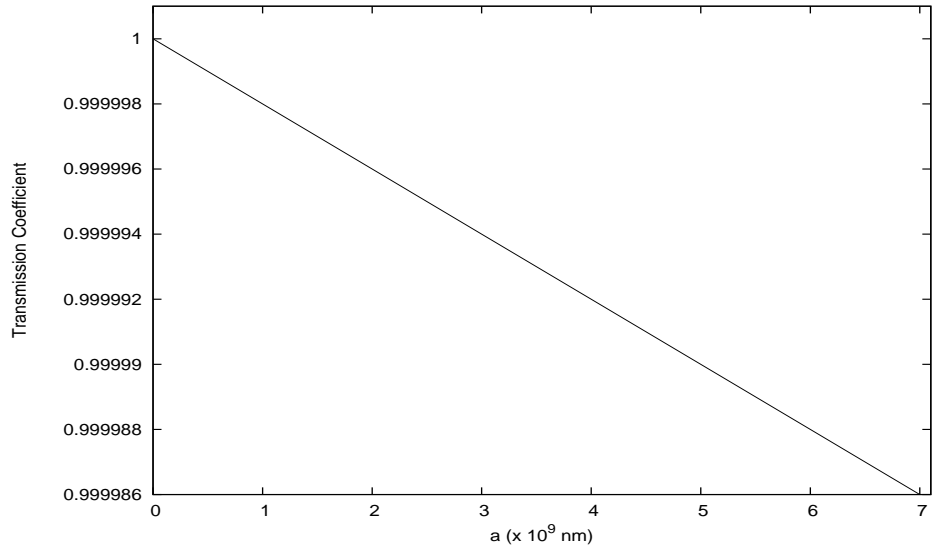}
\caption{The transmission probability versus the width of barrier
$a$.}
\end{figure}

\newpage

\begin{figure}
\centering
\includegraphics[height=3in, width=5in, angle=0]{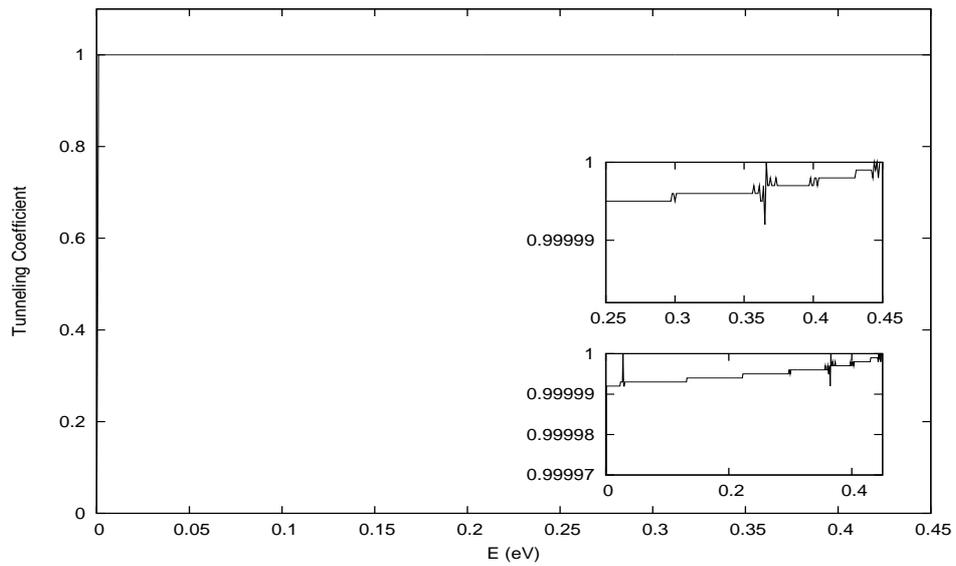}
\caption{The tunnelling coefficient for the triangular quantum
barrier versus $E$.}
\end{figure}

\end{document}